\documentclass [twocolumn,superscriptaddress,preprintnumbers,amsmath,amssymb,showpacs,prb]{revtex4}

\newcommand{\IMSS}{Muon Science Laboratory and Condensed Matter Research Center, Institute of Materials Structure Science, High Energy Accelerator Research Organization, Tsukuba, Ibaraki 305-0801, Japan}
\newcommand{\Sokendai}{Department of Materials Structure Science, The Graduate University for Advanced Studies (Sokendai),  Tsukuba, Ibaraki 305-0801, Japan}

\newcommand{\msr}{$\mu$SR\xspace}

\newcommand{\PRL}{Phys.~Rev. Lett.\xspace}
\newcommand{\PRB}{Phys.~Rev. B\xspace}

\newcommand{\fes}{FeS$_2$}
\usepackage{txfonts}
\usepackage{graphicx} 
\usepackage{dcolumn} 
\usepackage{bm} 
\usepackage{mediabb}
\usepackage{xspace}
\usepackage{ulem} 
\begin{document}

\title{Local Electronic Structure of Interstitial Hydrogen in Iron Disulfide}

\author{H.~Okabe}
\affiliation{\IMSS}\affiliation{\Sokendai}
\author{M.~Hiraishi}
\affiliation{\IMSS}
\author{S.~Takeshita}
\affiliation{\IMSS}
\author{A.~Koda}
\affiliation{\IMSS}\affiliation{\Sokendai}
\author{K.~M.~Kojima}
\affiliation{\IMSS}\affiliation{\Sokendai}
\author{R.~Kadono}\thanks{Corresponding author: ryosuke.kadono@kek.jp}
\affiliation{\IMSS}\affiliation{\Sokendai}

\begin{abstract}
The electronic structure of interstitial hydrogen in a compound semiconductor \fes\ (naturally $n$-type) is inferred from a muon study. An implanted muon (Mu, a pseudo-hydrogen) forms electronically different defect centers discerned by the hyperfine parameter ($\omega_{\rm hf}$). A body of evidence indicates that one muon is situated at the center of an iron-cornered tetrahedron with nearly isotropic $\omega_{\rm hf}$ (Mu$_{\rm p}$), and that the other exists as a diamagnetic state (Mu$_{\rm d}$, $\omega_{\rm hf}\simeq0$). Their response to thermal agitation indicates that the Mu$_{\rm d}$ center accompanies a shallow level (donor or acceptor) understood by effective mass model while the electronic structure of Mu$_{\rm p}$ center is more isolated from host than Mu$_{\rm d}$ to form a deeper donor level. These observations suggest that interstitial hydrogen also serves as an electronically active impurity in \fes. Based on earlier reports on the hydrogen diffusion in \fes, possibility of fast diffusion for Mu$_{\rm p}$ leading to formation of a complex defect state (Mu$^*_{\rm d}$, $T\le100$ K) or to motional narrowing state (Mu$^*_{\rm p}$, $T\ge150$ K) is also discussed.
\end{abstract}


\maketitle

\section{INTRODUCTION}
Iron disulfide (\fes), also known as the mineral pyrite or ``fool's gold", has significant scientific interest and technological applications. It was first explored as a photovoltaic semiconductor in the mid-1980s \cite{Ennaoui:93} and has attracted renewed attention in recent years \cite{Wadia:09a,Lin:09,Puthussery:11,Berry:12,Bi:11,Antonov:09,Hu:12,YZhang:12} as other thin-film absorber materials like amorphous silicon, CdTe, and Cu(In,Ga)Se$_2$ (CIGS) have gained commercial success.\cite{Bosio:11,Green:07} It is a promising optoelectronic material due to its suitable indirect band gap ($E_{\rm g}\simeq0.95$ eV) and high absorption coefficient ($>10^5$ cm$^{-1}$ at $E_{\rm g} \pm0.1$ eV), which opens up great potential for emerging renewable energy applications, including photovoltaics, photodetectors, and photoelectrochemical cells.\cite{Puthussery:11,Wang:12}  Interest in pyrite has also revived due to its low toxicity, virtually infinite elemental abundance, and low raw material cost.\cite{Wadia:09a,Lin:09,Puthussery:11,Berry:12,Bi:11,Antonov:09,Hu:12,YZhang:12,Wadia:09b}

Yet another interesting possibility for pyrite is its use as a dilute magnetic semiconductor for spintronics applications. It is now believed that high-temperature ferromagnetism in compound semiconductors reported previously is of extrinsic origin, resulting primarily from the precipitation of magnetic nanoparticles.\cite{Dietl:10}  Model calculations for \fes\ under the local density approximation (LDA) indicate that $t_{2g}$ orbitals in Fe are the primary component of the valence band maximum, whereas the conduction band minimum is dominated by Fe $e_g$ orbitals.\cite{Antonov:09,Wang:12} Incorporation of Co into pyrite at a concentration greater than 1\% results in percolative ferromagnetic order carried by the $e_g$ band.\cite{Guo:08} The narrow bandgap and high carrier concentration of pyrite may permit a stronger exchange interaction among local magnetic moments and, hence, a higher Curie temperature.  Furthermore, Fe is known for its stable high-spin state in most environments, suggesting that its magnetization is sensitive to point defects like vacancies or substituted impurities. 

The main obstacle to the development of pyrite as an optoelectronic material is its low open-circuit photovoltage ($V_{\rm oc}$), which is typically only $\sim$0.2 V. Traditionally, this has been attributed to surface defect states in \fes, its heterogeneous bandgap, and Fermi level pinning.\cite{DZhang:12,Yu:11} However, recent theoretical investigations suggest many different views, including one which suggests that sulfur vacancies are not the cause of these difficulties.\cite{Yu:11} Meanwhile, it has been known for decades that natural pyrite crystals often exhibit $n$-type conductivity of unknown origin with activation energies less than 0.01 eV.\cite{Schieck:90} There is circumstantial evidence that hydrogen is involved in this process.\cite{Ennaoui:93}  Moreover, electrochemical experiment suggests strikingly fast hydrogen diffusion in pyrite (corresponding diffusion coefficient $D_H\ge2\times10^{-6}$ cm$^2$/s, comparable to that in bcc metals at ambient condition),\cite{Wilhelm:83,Bungs:97} which is further enhanced after saturation of defects by hydrogen.\cite{Bungs:97} Considering that hydrogen is the most ubiquitous impurity, one may be naturally led to suspect interstitial hydrogen as the cause of these mysterious electrical activities in \fes.

It is well established that a positive muon ($\mu^+$) implanted into matter can be regarded as a light proton isotope in the sense that the local structure of a muon-electron system is virtually equivalent with that of hydrogen, except for a small correction ($\simeq0.4$\%) due to the difference in the reduced electron mass. While the light mass of muon ($\simeq m_{\rm p}/9$, with $m_{\rm p}$ being the proton mass) often leads to the isotope effect which is particularly distinctive in  diffusion at low temperatures where quantum tunneling process becomes dominant, muon also simulates hydrogen diffusion via classical over-barrier jump at high temperatures (which is demonstrated in a typical example of muon diffusion in iron \cite{Graf:80}).  Thus, muon in matter can be regarded as a pseudo-hydrogen. We propose the designation ``muogen" (Mu) as the appropriate elemental name, because the term ``muonium" exclusively refers to the neutral bound state of $\mu^+$ and $e^-$, analogous to atomic hydrogen. The electronic state of Mu can be spectroscopically identified via muon-electron hyperfine parameters using the muon spin rotation (\msr) technique with utmost sensitivity. 

Here, it is inferred from an implanted-muon study that there are four electronically inequivalent Mu states in \fes, i.e., two paramagnetic centers labeled Mu$_{\rm p}$ and Mu$^*_{\rm p}$, and two diamagnetic centers labeled Mu$_{\rm d}$ and Mu$^*_{\rm d}$.  
The magnitude of the hyperfine parameter [$\omega_{\rm hf}/2\pi\simeq411(40)$ MHz for Mu$_{\rm p}$ and $\omega_{\rm hf}=0$ for Mu$_{\rm d}$], combined with the Hartree potential calculation, suggest that Mu$_{\rm p}$ occupies an Fe-tetrahedron center corresponding to the S-S anti-bonding site. It is inferred from time-dependent muon polarization that Mu$_{\rm p}$ exhibits fast conversion to a diamagnetic state Mu$^*_{\rm d}$ ($\omega_{\rm hf}=0$, exhibiting depolarization due to spin/charge exchange interaction), which suggests a possibility of diffusion-limited trapping of  Mu$_{\rm p}$ to defects/impurities to form complex states. Mu$_{\rm d}$ is tentatively attributed to an effective mass-like shallow donor/acceptor state or a sulfhydryl-like SMu$^-$ complex that serves as donor by releasing an electron upon formation via the following process: S$_2^{2-}$ + Mu $\rightarrow$ S$^{2-}$ + SMu$^-$ + $e^-$. Their small ionization energy ($E_{\rm p} \simeq 10$ meV for Mu$_{\rm p}$ and $E_{\rm d} \le 1$ meV for Mu$_{\rm d}$) indicates that the electronic levels associated with these Mu centers are situated near (or merged to) the band edge. Meanwhile, the electronic state of Mu$^*_{\rm p}$ inferred from a positive frequency shift under a high transverse field is interpreted as Mu$_{\rm p}$ undergoing strong dynamical modulation of $\omega_{\rm hf}$ probably due to fast diffusion. These observations suggest that interstitial hydrogen also serves as an electronically active impurity in \fes.

\section{EXPERIMENT}
\begin{figure}[t]
 \centering
 \includegraphics[width=0.9\linewidth, clip]{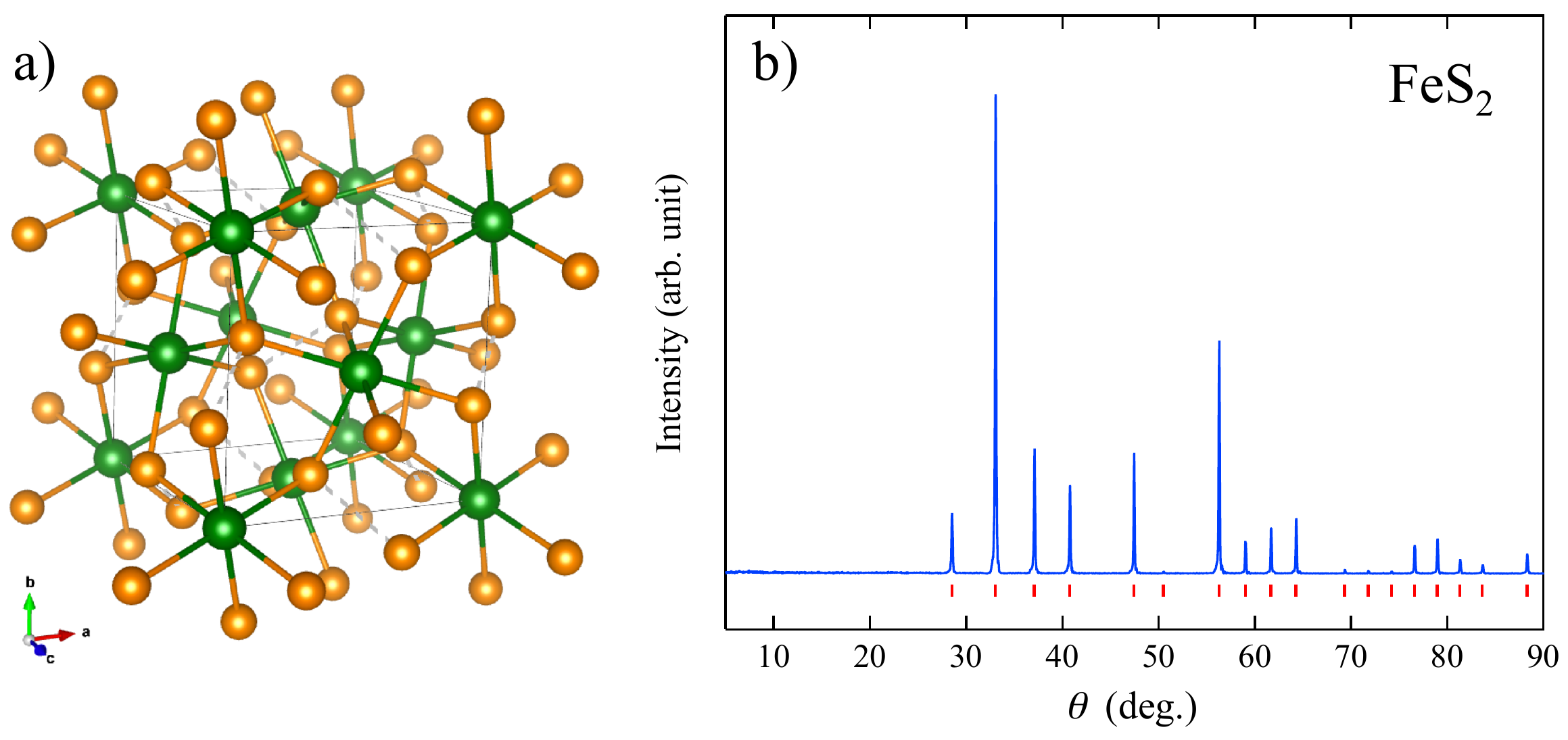}
 \caption{(Color online) a) Crystal structure of \fes, where green and brown balls represent Fe and S atoms. b) Powder x-ray diffraction spectrum obtained for the present \fes\ sample, where red lines indicate predicted peak positions.\cite{Brostigen:69}}
  \label{xrd}
\end{figure}

A single-crystalline ingot of natural pyrite (unknown origin) was sliced into slabs with planes normal to the [100] and [110] crystal axes for \msr\ measurements. A small portion of these slabs was used for powder x-ray diffraction (XRD) measurement and for bulk property characterization in order to investigate magnetic impurities and carrier concentration by uniform susceptibility ($\chi$, with magnetization measured under 1 T), resistivity, and Hall coefficient measurements. The crystal structure of \fes\ (shown in Fig.~\ref{xrd}a) belongs to a space group $Pa3$ (No.~205), consisting of FeS$_6$ octahedrons with S vertices forming dimers between them.  The powder XRD spectra in Fig.~\ref{xrd}b indicates that the sample was in a single phase with less that 1\% of impurities/defects.  

\begin{figure}[t]
 \centering
 \includegraphics[width=0.9\linewidth, clip]{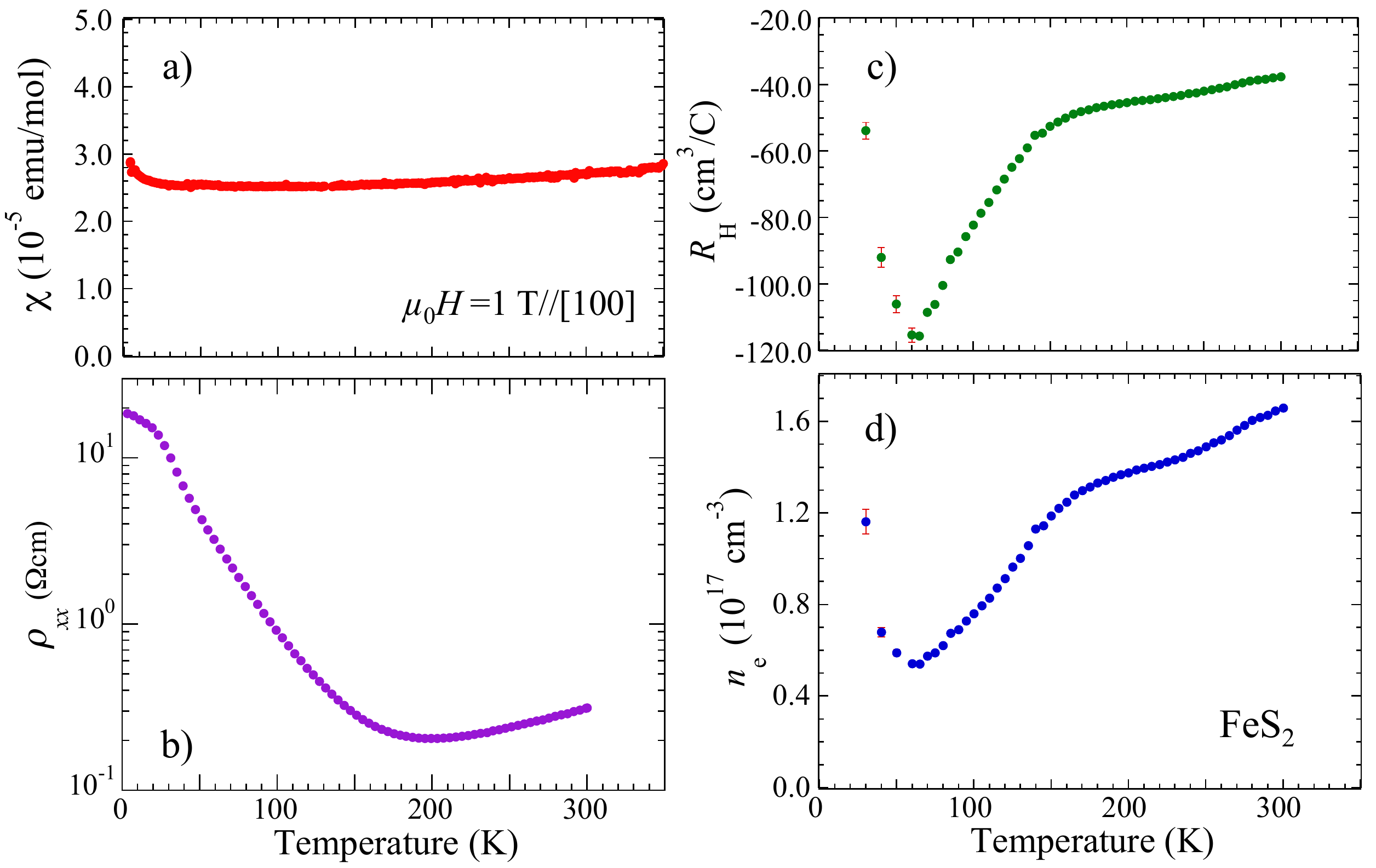}
 \caption{(Color online) Temperature dependence of bulk properties for the present \fes\ specimen. (a) Uniform magnetic susceptibility under a magnetic field of 1 T applied parallel to the [100] crystal axis, (b) electrical resistivity, c) Hall coefficient, and d) $n$-type carrier concentration evaluated from c).   }
  \label{chine}
\end{figure}

Regarding magnetic impurities, $\chi$ is almost completely independent of temperature, except for a slight enhancement below $\sim$20 K (see Fig.~\ref{chine}a). A curve-fit by the Curie-Weiss law for data below 50 K yields an effective moment $p_{\rm eff}=0.0125(3)\mu_B$, which corresponds to an atomic concentration of 1.11(3)$\times10^{20}$ cm$^{-3}$ for spin $S=1$ paramagnetic impurities (e.g., those associated with Fe vacancies, V$_{\rm Fe}$.\cite{Hu:12}) This paramagnetic defect center may be labeled ``X$_{\rm p}$."

Meanwhile, the negative sign of the Hall coefficient ($R_{\rm H}$, Fig.~\ref{chine}c) indicates that the residual carriers are dominated by $n$-type impurities.  The temperature dependence of $R_{\rm H}$, as well as that of electrical resistivity ($\rho_{xx}$, Fig.~\ref{chine}b) suggests that there are at least two species of unidentified donor centers with different activation energies whose origins are hereby labeled  ``X$_i$"  ($i=1,2$). They accompany donor levels $E_{{\rm X}i}$ with $|E_{\rm X1}|/k_B\simeq10^2$ K and $|E_{\rm X2}|/k_B\gg200$ K, yielding electronically active carriers of $n_e=$ 1--2$ \times10^{17}$ cm$^{-3}$ ($\simeq$ 4--8 atomic ppm). The relatively small $n_e$ compared with the concentration of X$_{\rm p}$ suggests that the paramagnetic electrons associated with X$_{\rm p}$  center are mostly localized up to ambient temperature.

Conventional $\mu$SR experiments were performed using the ARTEMIS spectrometer installed in the S1 area at J-PARC MUSE in order to measure the time-dependent $\mu$-$e$ decay asymmetry $A(t)$ under a zero field (ZF), longitudinal field (LF), and weak transverse field (TF). A 100\% spin-polarized pulsed beam (FWHM $\simeq100$ ns) of positive muons with a momentum of 27 MeV/c was used to irradiate a single-crystalline \fes\ slab loaded on a He-flow cryostat to monitor $A(t)$ over a temperature range from 3.5 to 300 K. Additional measurements were conducted using the NuTime spectrometer on the M15 beamline at TRIUMF, Canada, in order to measure the \msr\ frequency shift under a high transverse field of 6 T. 

The time evolution of muon polarization for the muonium-like paramagnetic state is described by that of the spin-triplet ($F=1$) and spin-singlet ($F=0$) states. The muon spin polarization under LF (applied to the $\hat{z}$ direction parallel to the initial spin polarization) is described by 
\begin{equation}
P_z^{\rm p}(t;b)= \frac{1}{2(1+b^2)}\left[(1+2b^2)+\cos\omega_b t\right]\simeq\frac{1+2b^2}{2(1+b^2)},\label{lfd}
\end{equation} 
where $b=\omega_\mu/\omega_c$, $\omega_\mu=\gamma_\mu B$, $\omega_c=\omega_{\rm hf}\gamma_\mu/(\gamma_\mu-\gamma_e)$, $\gamma_\mu$  ($=2\pi\times 135.53$ MHz/T), $\gamma_e$ ($=2\pi\times 28024.21$ MHz/T) are the muon and electron gyromagnetic ratio, respectively, and $\omega_b=\omega_{\rm hf}(1+b^2)^{1/2}$ is the muon spin precession frequency for the $F=0$ state. Because $\omega_b$ usually exceeds the limit determined by the experimental time resolution (the Nyquist frequency for the time resolution of 100 ns at J-PARC MUSE is 5 MHz), the second term cannot be resolved (i.e., averaged to zero). The residual polarization ($=1/2$ for $B=0$) corresponds to the $F=1$ state.
Also disregarding the unresolved $F=0$ state, the time evolution under a TF ($\omega_\mu\ll \omega_{\rm hf}$) is approximately given by that for the $F=1$ state,
$P_x^{\rm p}(t)\simeq \frac{1}{2}\cos\omega_{\rm p} t$,
where $\omega_{\rm p}= (\gamma_\mu-\gamma_e)B/2\simeq\gamma_eB/2$.  

Meanwhile, the response of the diamagnetic muon (Mu$^+$ or Mu$^-$ state) due to the external field is described by $P_z^{\rm d}(t)=1$ for LF and $P_x^{\rm d}(t)=\cos\omega_\mu t$ for TF when the spin fluctuation of nearby electrons is negligible. Note that $\omega_{\rm p}$ for the paramagnetic state is nearly a hundred times greater than $\omega_\mu$ for the diamagnetic state.
We also point out that the modulation of $P_{z,x}^{\rm d}(t)$ due to random local fields from nuclear magnetic moments is negligible for \fes\ because of small natural abundance of isotopes with non-zero spin nuclei [$^{57}$Fe (2.14\%) and $^{33}$S (0.75\%)]. 

\section{RESULTS and DISCUSSION}
\subsection{Electronic structure and dynamics of Mu below $\sim$100 K}
\begin{figure}[t]
 \centering
 \includegraphics[width=0.85\linewidth, clip]{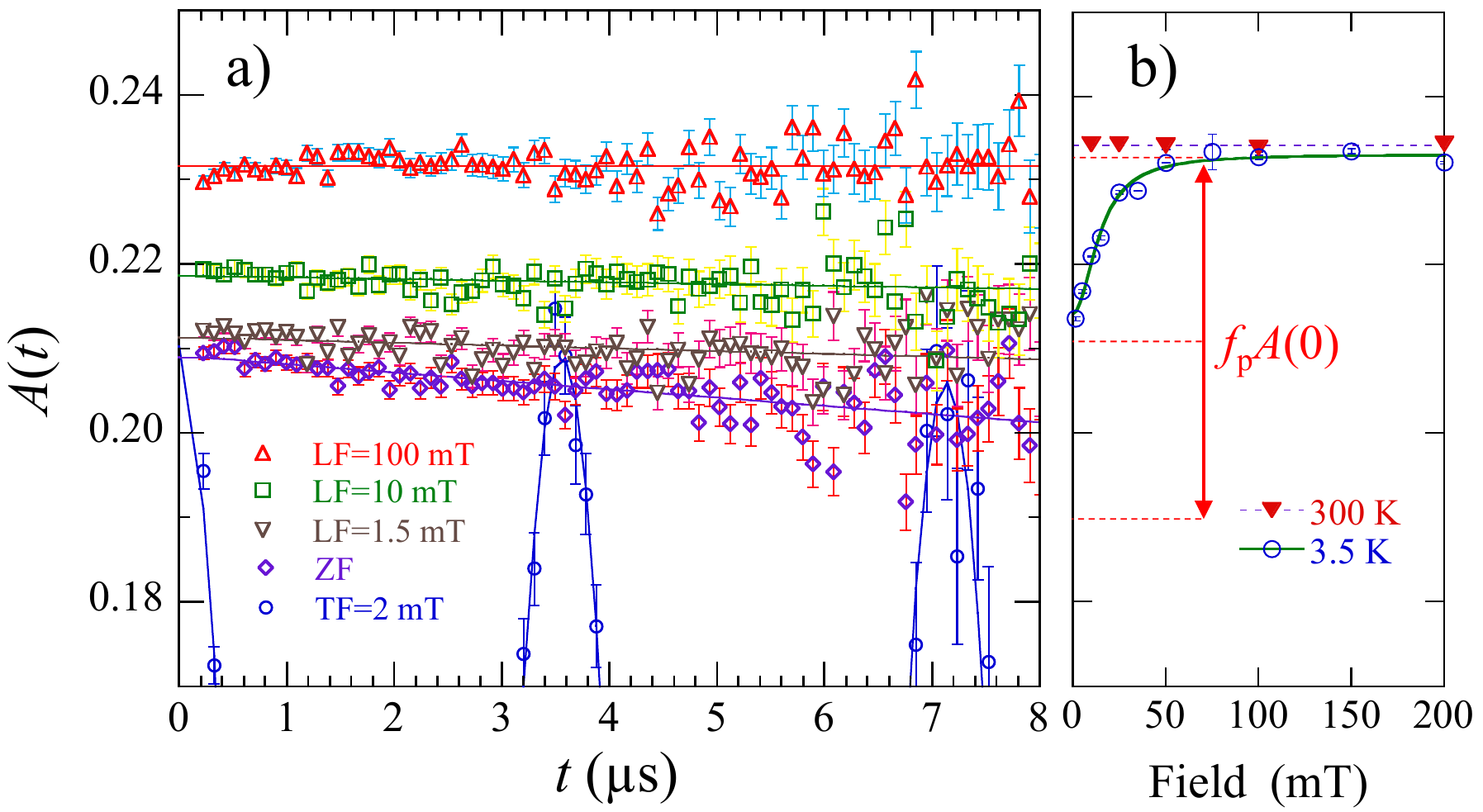}
 \caption{(Color online) a) Typical examples of \msr\ spectra observed at 3.5 K for the \fes\ sample with [100] orientation under a longitudinal field of zero (ZF), 1.5, 10, and 100 mT (applied parallel to the [100] axis). Only a portion is visible for the spectrum under a $2$ mT TF. b) LF dependence of the initial asymmetry $A(0)$ at 3.5 K. Solid line indicates a result of curve fit by Eqs.~(\ref{lfd}) and (\ref{lf}) in the text, where the LF dependence of $A(0)$ at 300 K (dashed curve) was used as a normalization to cancel the weakly field-dependent instrumental asymmetry. Arrowed bar indicates the partial asymmetry for Mu$_{\rm p}$.}
  \label{tsp4k}
\end{figure}

Figure \ref{tsp4k} shows the \msr\ spectra at 3.5 K observed under various conditions for the external magnetic field, which is magnified along the vertical axis to improve the visibility of $A(t)$. The increase of $A(0)$ from 0.210(6) to 0.230(6) upon an LF increase from zero to 100 mT clearly indicates that a fraction of muons form a paramagnetic state (which is tentatively labeled as Mu$_{\rm p}$). The spectrum under TF = 2 mT is perfectly reproduced by $P_x^{\rm d}(t)$, indicating that the rest of the precession signal is attributed to the diamagnetic state, which is labeled as Mu$_{\rm d}$.
 All \msr\ spectra are then expected to be reproduced by recursive functions
\begin{eqnarray}
A_{\rm LF}(t)&\simeq& A(0)[f_{\rm p}P_z^{\rm p}(t;u)+f_{\rm d}e^{-t/T_1}+f_{\rm b}]\label{lf}\\
A_{\rm TF}(t)&\simeq& A(0)[\frac{f_{\rm p}}{2}P_x^{\rm p}(t)+(f_{\rm d}e^{-t/T_1} +f_{\rm b})\cos\omega_\mu t ]\label{atf}
\end{eqnarray}
for LF and TF, respectively, 
where $f_{\rm p}$, $f_{\rm d}$, and $f_{\rm b}$  ($f_{\rm p}+f_{\rm d}+f_{\rm b}=1$) are respectively the fractional yield of Mu$_{\rm p}$, Mu$^\pm$, and the background (typically $f_{\rm b}\sim$0.1),  and $1/T_1$ is the depolarization rate due to electron spin fluctuations. A curve-fit analysis including the field dependence for the LF-\msr spectra yields $\omega_{\rm hf}/2\pi=411(40)$ MHz and $f_{\rm p}=0.163(8)$ (see Fig.~\ref{tsp4k}b). Similar measurements with LF applied parallel to the [110] axis yields $\omega_{\rm hf}/2\pi=417(45)$ MHz, suggesting that the hyperfine interaction has the least anisotropy.

Here, the first term in Eq.~(\ref{atf}) predicts a relatively fast spin precession of $P_x^{\rm p}(t)$ with a frequency $\omega_{\rm p}/2\pi\simeq28$ MHz at 2 mT.  As ilustrated in Fig.~\ref{tsp4k}b, this should lead to a reduction in the initial asymmetry for the TF-\msr\ spectra by an amount $A_{\rm LF}(0)-A_{\rm TF}(0)=f_{\rm p}A(0)\simeq0.038$, as the term $P_x^{\rm p}(t)$ would be reduced to zero due to the limited time resolution. The absence of such reduction in the actual spectrum in Fig.~\ref{tsp4k} [as extra asymmetry $f_{\rm p}A(0)/2$ remains] indicates occurrence of irreversible process from Mu$_{\rm p}$ to a diamagnetic state Mu$^*_{\rm d}$ (not necessarily identical with Mu$_{\rm d}$) with a rate faster than $\omega_{\rm p}$, which leads to the increase of $A_{\rm TF}(0)$ by $f_{\rm p}A(0)/2\simeq0.019$.  Considering the fast diffusion inferred for hydrogen in \fes, we tentatively attribute the microscopic origin of the process to the diffusion-limited formation of a complex state between Mu$_{\rm p}$ and defects/impurities including the X$_{\rm p}$ center. The disappearance of Mu$_{\rm p}$ with increasing temperature towards $\sim$100 K is then attributed to the increase of conversion rate to Mu$^*_{\rm d}$ due to thermally enhanced muon diffusion. 

\begin{figure}[t]
 \centering
 \includegraphics[width=0.9\linewidth, clip]{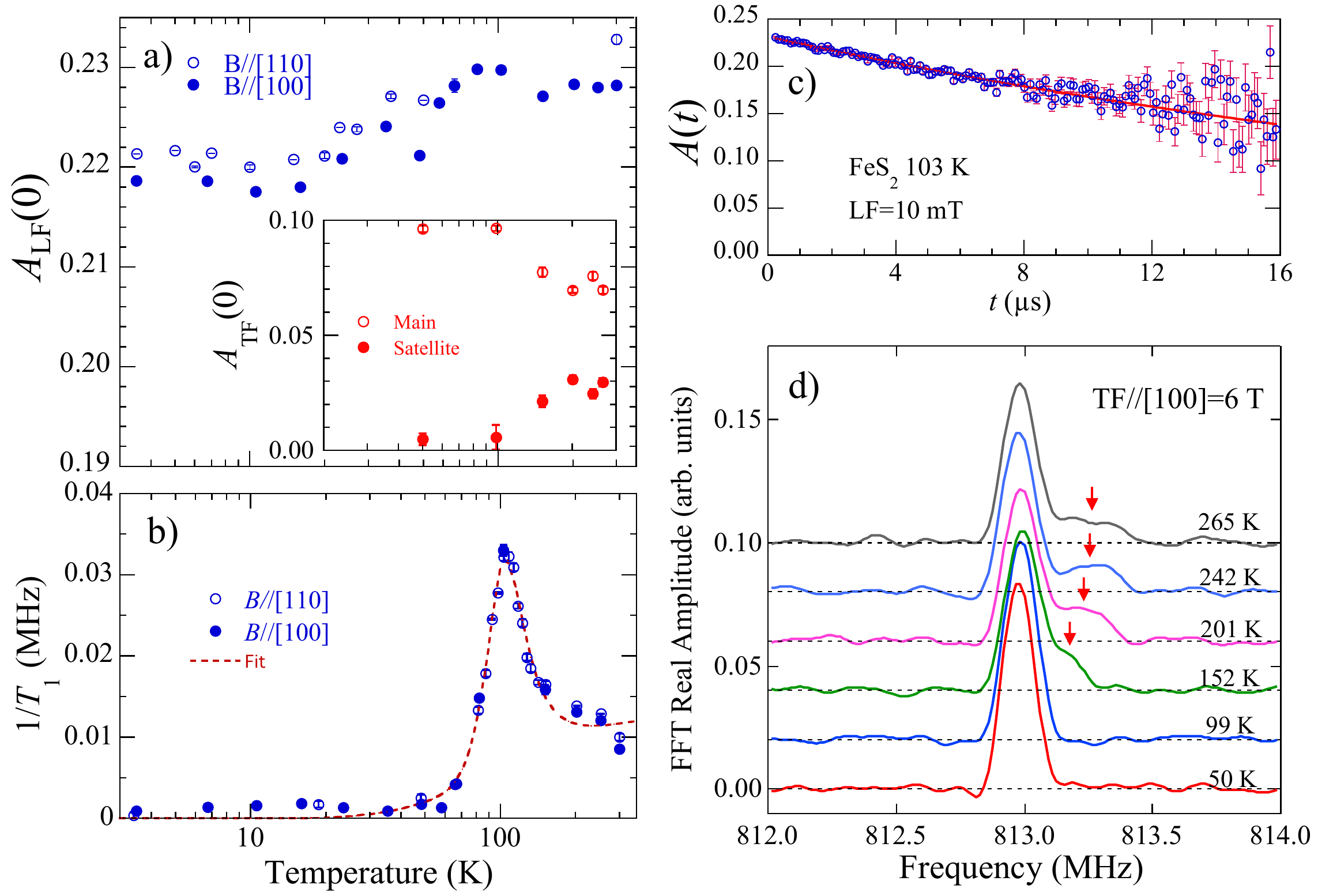}
 \caption{(Color online) Temperature dependence of a) the initial asymmetry and b) longitudinal depolarization rate under LF = 10 mT in \fes. The dashed curve in b) shows result of the least-square fit using the Redfield model (see text). c) A typical example of an LF-\msr\ time spectrum observed at 103 K. d) Fast Fourier transform of the \msr\ time spectra measured under TF = 6 T, where the partial asymmetry of the main and satellite (arrow markers in d) precession signals [$f_{\rm d}A_{\rm TF}(0)$ and $f_{\rm p}A_{\rm TF}(0)$] are shown in the inset of a). }
  \label{t1fft}
\end{figure}

As shown in Fig.~\ref{t1fft}a, $A_{\rm LF}(0)$ exhibits a gradual increase with increasing temperature, reaching full asymmetry (corresponding to the initial spin polarization of 100\%) above $T^*\simeq80$ K. The concomitant change of $1/T_1$ with a peak around $\sim$100 K (Figs.~\ref{t1fft}b and \ref{t1fft}c) is interpreted as a monotonic spin/charge exchange interaction enhancement between Mu$^*_{\rm d}$ and thermally promoted carriers from the X$_i$ donors. Note that such a depolarization immediately implies presence of a Mu$_{\rm p}$-like intermediate state with a life time greater than $\omega_{\rm hf}^{-1}$ in the spin/charge exchange process.  The temperature dependence, including the peak structure corresponding to the ``$T_1$-minimum," is perfectly reproduced by the Redfield model, 
\begin{equation}
1/T_1\simeq\frac{2\delta_{\rm shf}^2\nu(T)}{\omega_\mu^2+\nu^2(T)} +\lambda_{\rm p}(T), \label{rfd}
\end{equation}
combined with the fluctuation rate controlled by thermal excitation,
$\nu(T)=\nu_0\exp(-E_{\rm a}/k_BT)$, where $\delta_{\rm shf}$ is the super-hyperfine coupling between muon and carriers via the electron bound to the Mu$^*_{\rm d}$ complex, $\nu_0$ is the exchange rate, $E_{\rm a}$ is the activation energy, and $\lambda_{\rm p}(T)$ is the additional contribution emerging when $T>T^*$.
Provided that the influence of acceptor impurities is negligible, we can assume $\nu_0\simeq\sigma_\mu v_e(N_DN_C/2)^{1/2}$ and $E_{\rm a}\simeq E_{{\rm X}i}/2$, where $\sigma_\mu$ is the cross section for Mu-carrier interaction, $v_e$ is the carrier velocity, $N_D$ and $N_C$ are the respective density of states for the X$_i$ donors and the conduction band bottom. A least-square fit of the data using Eq.~(\ref{rfd}) with the further assumption that $\lambda_{\rm p}(T)=\lambda_0\exp(-E_{\rm p}/k_BT)$ yields $\delta_{\rm shf}=0.473(17)$ MHz, $\nu_0=5.0(1.8)\times10^9$ s$^{-1}$, $E_a=57(3)$ meV, $\lambda_0=0.0153(16)$ MHz, and $E_{\rm p}/k_B=102(21)$ K.  The correspondence between $E_{\rm p}/k_B$ and $T^*$ suggests that $E_{\rm p}$ is related to the stability of the Mu$^*_{\rm d}$ complex state against ionization or charge exchange interaction, i.e., ${\rm Mu^*_d}\leftrightarrow {\rm Mu^{*+}_d} + e^-$, where Mu$^{*+}_{\rm d}$ involves the Mu$_{\rm p}$-like state. The magnitude of $E_{\rm a}$ suggests that the spin/charge exchange process for $T\ge 100$ K is dominated by carriers promoted from the X$_2$ donor. 

\subsection{Spin/charge dynamics of Mu above $\sim$150 K}
In raising temperature above $\sim$100 K, another diamagnetic state discernible only by muon Knight shift under a high TF ($B=6$ T) is observed. As shown in Fig.~\ref{t1fft}d, a satellite signal develops at the higher frequency side of the central peak [$\omega_{\mu}=812.98(1)$ MHz]. Curve-fit analysis in the time domain by a function $A(t)=A_{\rm TF}(0)[f_{\rm d}\cos(\omega_{\mu}t+\phi)+f_{\rm p}\cos(\omega'_{\mu}t+\phi)]$ ($f_{\rm b}=0$ for these measurements) provides the fractional yield of the satellite signal $f_{\rm p}$ and the frequency shift $K_\mu=(\omega'_{\mu}-\omega_{\mu})/\omega_{\mu}$. The mean values of the data for 200--265 K are obtained as $f_p=0.283(19)$ and $K_\mu=+279(19)$ ppm, where the latter is orders of magnitude greater than that expected for the chemical shift ($<10$ ppm).

While the single satellite signal is apparently inconsistent with the case for a shallow donor-like state (which usually accompanies two satellite signals at $\omega_\mu\pm\omega_{\rm hf}/2$), the temperature dependence of $K_\mu$ disfavors the possibility of ascribing the signal to local paramagnetic defects/impurities of extrinsic origin (for which we expect $K_\mu\propto 1/T$). It is also noticeable that $K_\mu$ is not proportional to $\chi$, as $\chi$ does not exhibit any steplike change around 150--200 K (see Fig.~\ref{chine}a). Thus, the signal is presumed to be associated with the electronic state involving muon as a defect center (which we call Mu$^*_{\rm p}$ henceforth). We discuss two mechanisms that may account for the strong modulation of hyperfine parameters, namely, i) fast spin/charge exchange of a stationary Mu$_{\rm p}$-like state with thermally promoted carriers, and ii) fast spin flip due to diffusion of the Mu$_{\rm p}$-like state. 

In the case i), according to the model of spin/charge dynamics for Mu in heavily $n$-doped Si, the muon spin precession frequency $\omega'_{\mu}$ in the limit of fast spin/charge exchange is determined by the mean hyperfine field exerted from the 1$s$ orbital electron that is polarized (obeying the Curie-Weiss law) by an external field.\cite{Chow:00} For the isotropic hyperfine parameter, the corresponding shift is evaluated as
\begin{equation}
K_\mu=\frac{\omega'_{\mu}-\omega_{\mu}}{\omega_{\mu}}=\frac{h\gamma_e\omega_{\rm hf}^*}{8\pi\gamma_\mu k_BT}\:,\label{Kmu}
\end{equation}
where $\omega_{\rm hf}^*$ is the effective hyperfine parameter for Mu$^*_{\rm p}$ which is reduced from $\omega_{\rm hf}$ by charge screening.
Note in Eq.~(\ref{Kmu}) that the sign of the shift implies the sign of $\omega_{\rm hf}^*$ (which can be either positive or negative, depending on the local electronic structure).  
Using the observed value for $K_\mu$, $\omega_{\rm hf}^*/2\pi$ is coarsely estimated to be 25--33 MHz for 200--265 K with a positive sign. While the sign is consistent with the interpretation that the signal comes from the Mu$_{\rm p}$ center, the magnitude of $\omega_{\rm hf}^*$ is considerably smaller than that expected for the relevant carrier concentration ($n_e\sim$10$^{17}$ cm$^{-3}$).\cite{Chow:00} The weak $T$ dependence of the shift is also against this interpretation, because Eq.~(\ref{Kmu}) predicts $K_\mu\propto1/T$.

Concerning the case ii), we note that the jumping frequency ($\nu$) for hydrogen suggested from the reported diffusion coefficient is extremely high: $\nu\simeq zD_H/d^2\ge1.7\times 10^{11}$ s$^{-1}$ at ambient temperature, where we assumed $z=6$ and $d=a/2$ for a presumed interstitial site (see below) with $a$ (= 0.5428 nm) being the lattice constant for the cubic unit cell of \fes. The influence of fast diffusion is similar to the case i) as long as it induces spin relaxation of orbital electron due to the modulation of the hyperfine interaction associated with jump from one site to another.  Assuming that the Mu$^*_{\rm p}$ also undergoes diffusion comparable to hydrogen, the ratio $\gamma\equiv4\nu/\omega_{\rm hf}$ is about $10^2$ in order of magnitude. According to the theories of spin dynamics,\cite{Nosov:62,Ivanter:68,Patterson:88} such a fast spin flip leads to strong reduction of the effective hyperfine parameter by a factor $\sim\gamma^{-1}$ (i.e., the motional narrowing), which is qualitatively in line with the present result.

A plausible scenario emerging from these discussions is that Mu$^*_{\rm p}$ is identical with the Mu$_{\rm p}$ center undergoing fast diffusion, where the diffusion leads to Mu$^*_{\rm d}$ complex formation with defects/impurities below $\sim$100 K and to strong modulation of electronic structure at high temperatures.  The temperature dependence of the yield (see Fig.~\ref{t1fft}a inset)  suggests that Mu$^*_{\rm p}$ may partially originate from muon released from the Mu$^*_{\rm d}$ complex state by thermal agitation (detrapping from the defect/impurity center).

\subsection{Electronic structure of Mu/H}
\begin{figure}[t]
 \centering
 \includegraphics[width=0.9\linewidth, clip]{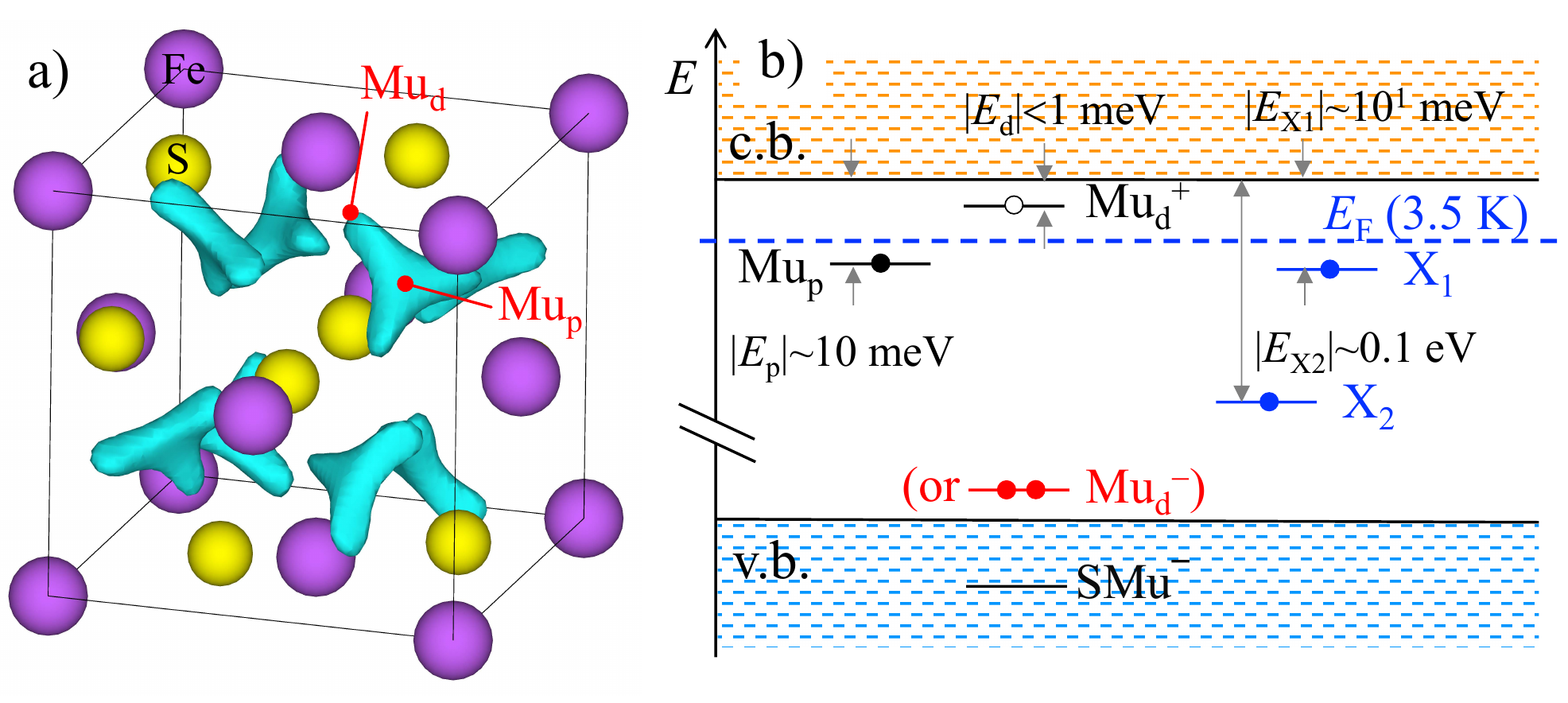}
 \caption{(Color online) a) Atomic configuration of \fes\ and possible candidate sites for two Mu states, where isosurfaces of the Hartree potential (0.3 eV above the $-11.387$ eV potential minimum) are displayed by blue-hatched areas. b) Schematic energy diagram of the electronic states associated with two interstitial Mu centers, Mu$_{\rm p}$ and Mu$_{\rm d}$, and the electronic levels of X$_i$ donors in naturally-occurring $n$-type \fes.
The Fermi level at 3.5 K corresponds to the lowest temperature for the present experiment to explain the charge states of these atomic defect centers.}   
  \label{enedia}
\end{figure}

Now, let us discuss the possible local structure of these Mu-related centers. As shown in Fig.~\ref{enedia}a, it is suggested from our preliminary calculation [using the Vienna {\it ab initio} Simulation Package (VASP)]~\cite{VASP} that the Hartree potential for the interstitial Mu$^+$ exhibits minima around the center of an Fe-cornered tetrahedron with lobes extending along trigonal directions perpendicular to the S-S bond axis. This naturally leads to the speculation that the position near the Fe-tetrahedron center may correspond to the S-S anti-bonding (AB) site.  According to our empirical rule that the muon (hydrogen) at such high-symmetry sites tends to form isolated defect centers with an isotropic hyperfine parameter, Mu$_{\rm p}$ may be attributed to the AB site muon. However, it must be noted that the electronic structure is not understood by the simple effective mass model because it predicts orders of magnitude smaller $\omega_{\rm hf}$. Adopting known values for the effective mass ($m^*/m_e\simeq0.45$ Ref.\onlinecite{Tsay:93}) and dielectric constant ($\epsilon'\simeq10.9$ Ref.\onlinecite{Husk:78}), we have $\omega_{\rm hf}\simeq \omega_{\rm vac}[m^*/(m_e\epsilon')]^3\simeq2\pi\times0.314$ MHz, where $\omega_{\rm vac}/2\pi=4463$ MHz is the hyperfine parameter for muonium in vacuum and $m^*/(m_e\epsilon')$ is the Bohr radius scaling factor. In this regard, the extremely low carrier promotion energy associated with the Mu$_{\rm d}$ center suggests that Mu$_{\rm d}$ accompanies the shallow donor state described by the effective mass model.  

Meanwhile, it is tempting to speculate on another possibility for the Mu$_{\rm d}$ center, which is similar to the case in rutile TiO$_2$, where the formation of a OMu$^{-}$ complex state is suggested at low temperatures.\cite{Shimomura:15,Vilao:15} The OMu$^{-}$ complex accompanies a loosely bound electron in the Ti $t_{2g}$ orbital (which is stable only below $\sim$10 K), comprising the ``large polaron" state with extremely shallow donor level ($\simeq1$ meV). Its local electronic structure suggests that Mu (and hence H as well) is prone to the OMu bound state via the lone-pair electrons of the O$^{2-}$ ligand that coordinates Ti by $sp^2$ hybridized orbitals, thereby promoting the reaction O$^{2-}$ + Mu $\rightarrow$ OMu$^{-}$ + $e^{-}$ (Ti$^{3+}$). Considering the similarity of the local electronic structure between \fes\ and TiO$_2$, the formation of a sulfhydryl-like SMu$^-$ complex is reasonably expected. The absence of the large-polaronic state even at 3.5 K then suggests the relatively large bandwidth of the 3$d$ $e_g$ orbitals that comprise the bottom of the conduction band.

It is further speculated that the trigonal lobes found for the Hartree potential minima may correspond to sites for the SMu$^-$ complex. This also leads to an estimated relative density of states to be about 3 vs.~1 for Mu$_{\rm d}$ vs.~Mu$_{\rm p}$, which is in semi-quantitative agreement with the experimental observations of their relative yields, $f_{\rm d}$ vs.~$f_{\rm p}$.  In this sense, it would not be necessary to presume correlation between Mu$^*_{\rm d}$ complex and Mu$^*_{\rm p}$ (=Mu$_{\rm p}$ under fast diffusion), because the yield is interpreted as a result of annealing to the ground state at ambient temperature. 

Finally, the energetics of the Mu-related centers is summarized in Fig.~\ref{enedia}b. One can infer from the spin/charge dynamics that the electronic level associated with Mu$_{\rm p}$ is situated at $E_{\rm p}\simeq k_BT^*\sim$10 meV below the conduction band, which seems shallow enough to be an electrically active impurity at ambient temperature. [It is tempting to attribute the X$_1$ donor to hydrogen (the H$_{\rm p}$  center) upon observing the coincidence between $E_{\rm p}$ and $E_{\rm X1}$.] The Mu$_{\rm d}$ center directly serves as a donor by forming either an effective mass-like shallow level or a putative SMu$^-$ complex, while the bonding levels associated with the latter are situated deep in the valence band.  The possibility of associating Mu$_{\rm d}$ as a shallow acceptor center (Mu$^-_{\rm d}$ at 3.5 K) also remains, although the local electronic structure is unclear.  It would be also worth noting that the Mu$^*_{\rm d}$ complex state might correspond to the muon trapped to iron vacancy, Mu$^+_{\rm Fe}$, considering the possibility of ascribing X$_{\rm p}$ center to V$_{\rm Fe}$.  The complex state may also serve as acceptor via the process, Mu$_{\rm p}$ + V$^{2+}_{\rm Fe} \rightarrow$ Mu$^+_{\rm Fe} + h^+$, which suggests the possibility that hydrogen compensation of V$_{\rm Fe}$ as origin of $p$-type doping in \fes.  Detailed theoretical analysis of hydrogen-related defects based on an {\it ab initio}-type calculation is in due for further understanding for the role of hydrogen in \fes.

\section*{ACKNOWLEDGMENTS}
We would like to thank the staff of KEK-MSL and TRIUMF for their technical support during the $\mu$SR experiments. This work was partially supported by the KEK-MSL Inter-University Research Program (Proposal No. 2016B0011) and by the MEXT Elements Strategy Initiative to Form Core Research Center for Electron Materials. We also acknowledge CROSS-Tokai for the use of MPMS and PPMS in their user laboratories, and H. Lee for VASP calculations under the support of KEK Large Scale Simulation Program No.~16-17.


\end{document}